\documentclass[pdflatex,sn-nature]{sn-jnl}

\usepackage{graphicx}%
\usepackage{multirow}%
\usepackage{amsmath,amssymb,amsfonts}%
\usepackage{amsthm}%
\usepackage{mathrsfs}%
\usepackage[title]{appendix}%
\usepackage{xcolor}%
\usepackage{textcomp}%
\usepackage{tabularx}
\usepackage{manyfoot}%
\usepackage{booktabs}%
\usepackage{algorithm}%
\usepackage{algorithmicx}%
\usepackage{algpseudocode}%
\usepackage{listings}%
\usepackage{lineno}
\usepackage{siunitx}
\usepackage{amsmath}
\usepackage{booktabs} 
\usepackage{booktabs}   
\usepackage{rotating}   
\usepackage{caption}    

\usepackage{threeparttable}
\usepackage[T1]{fontenc}   
\usepackage{lmodern}       

\usepackage{mathrsfs}
\DeclareFontFamily{U}{rsfs}{}
\DeclareFontShape{U}{rsfs}{m}{n}{<->rsfs10}{}
\DeclareFontShape{U}{rsfs}{m}{sl}{<->rsfsl10}{}
\DeclareFontShape{U}{rsfs}{bx}{n}{<->rsfs10}{}
\DeclareFontShape{OMS}{cmsy}{m}{n}{<-> s*[1.0] cmsy10}{}

\usepackage{array}      
\usepackage{booktabs}   
\usepackage{rotating}   
\newcolumntype{C}[1]{>{\centering\arraybackslash}p{#1}}
\renewcommand\thefootnote{\fnsymbol{footnote}}
\newcommand{\equalcontrib}{\textsuperscript{\dag}}
\newcommand{\corrauthor}{\textsuperscript{*}}
\usepackage[english]{babel}
\usepackage{amsmath}
\usepackage{physics}
\usepackage{hyperref}
\usepackage{multicol}
\usepackage{bbold}
\usepackage{hyperref}

\theoremstyle{plain}
\theoremstyle{plain}%

\theoremstyle{plain}%

\title{Neural Network Architectures for Scalable Quantum State Tomography: Benchmarking and Memristor-Based Acceleration}

\author[1]{Erbing Hua\equalcontrib\corrauthor}
\author[1]{Steven van Ommen\equalcontrib}
\author[1,2]{King Yiu Yu\equalcontrib}
\author[1]{Jim van Leeuven}
\author[1]{Rajendra Bishnoi}
\author[1]{Heba Abunahla}
\author[1,2]{Salahuddin Nur}
\author[1,2]{Sebastian Feld}
\author[1,2]{Ryoichi Ishihara\corrauthor}

\affil[1]{\hspace{0.5em}Department of Quantum and Computer Engineering, Delft University of Technology, Delft, The Netherlands}

\affil[2]{\hspace{0.5em}QuTech, Delft University of Technology, Delft, The Netherlands}

\abstract{
Quantum State Tomography (QST) is essential for characterizing and validating quantum systems, but its practical use is severely limited by the exponential growth of the Hilbert space and the number of measurements required for informational completeness. 
Many prior claims of performance have relied on architectural assumptions rather than systematic validation. 
We benchmark several neural network architectures to determine which scale effectively with qubit number and which fail to maintain high fidelity as system size increases.
To address this, we perform a comprehensive benchmarking of diverse neural architectures across two quantum measurement strategies to evaluate their effectiveness in reconstructing both pure and mixed quantum states. 
Our results reveal that CNN and CGAN scale more robustly and achieve the highest fidelities while Spiking Variational Autoencoder (SVAE) demonstrates moderate fidelity performance, making them strong candidates for embedded, low-power hardware implementations.
Recognizing that practical quantum diagnostics will require embedded, energy-efficient computation, we also discussed how memristor-based Computation-in-Memory (CiM) platforms can accelerate these models in hardware, mitigating memory bottlenecks and reducing energy consumption to enable scalable \textit{in-situ} QST. This work identifies which architectures scale favorably for future quantum systems and lays the groundwork for quantum–classical co-design that is both computationally and physically scalable.
}


\usepackage{etoolbox}
\AtBeginDocument{
  \geometry{top=2.5cm, bottom=2.5cm, left=2.5cm, right=3.0cm}
}

\begin{document}
\pagestyle{plain}

\maketitle

\begingroup
\renewcommand\thefootnote{}\footnotetext{\equalcontrib\ Equal contribution.}
\renewcommand\thefootnote{}\footnotetext{\corrauthor\ Corresponding author: \texttt{e.hua@tudelft.nl, r.ishihara@tudelft.nl}}
\endgroup

\section*{Introduction}

Quantum computing represents a groundbreaking paradigm that promises to redefine the boundaries of information processing. It leverages quantum superposition and entanglement to solve classically intractable problems in cryptography, simulation, and optimization~\cite{nielsen2010quantum,shor1999polynomial,farhi2014quantum,lloyd1996universal}. Therefore, understanding the quantum system is necessary to form a critical operational foundation for calibration, error detection, benchmarking, and validation of quantum devices and algorithms with QST from measured data~\cite{paris2004quantum}. However, QST faces a central challenge: exponential scaling in both the Hilbert space ($2^N$ for $N$ qubits) and required measurement bases ($4^N$), which hampers its practicality for large-scale quantum systems~\cite{gross2010quantum,cramer2010efficient}. 

To overcome this bottleneck, a complementary path has emerged: leveraging machine learning to reduce measurement or computational overhead. Recent advances in applying artificial intelligence (AI), particularly neural networks~\cite{melnikov2018active}, to QST have demonstrated strong potential for mitigating the curse of exponential dimensionality, by learning to reconstruct quantum states from fewer, noisy, or incomplete measurements~\cite{Torlai2018, carrasquilla2019reconstructing}. Numerous recent studies have demonstrated the effectiveness of various neural network architectures, including Convolutional Neural Networks (CNN)~\cite{schmale2022efficient,lohani2020machine,ma2024neural}, Fully Connected Networks (FCN), Recurrent Neural Networks (RNN)~\cite{morawetz2021u}, Restricted Boltzmann Machines (RBM)~\cite{neville2017classical}, Conditional Generative Adversarial Networks (CGAN)~\cite{Ahmed2021}, Transformers~\cite{ma2023attention, ma2025learning}, and Variational Autoencoders (VAE)~\cite{ rocchetto2018learning,chen2021reconstructing}. While recent studies demonstrate that neural networks–based QST can be effective for small numbers of qubits, scaling these approaches to larger, practical quantum systems remains a challenge. This scalability demands not only algorithmic efficiency but also energy-efficient hardware support. However, current software-centric methods rarely address these hardware constraints. Addressing these critical limitations necessitates a shift toward hardware-aware neural network architectures. Conventional von Neumann computing architectures, characterized by separated memory and processing units, are severely limited by the \textit{memory wall} problem, resulting from substantial data transfer bottlenecks that constrain computational efficiency and scalability \cite{backus1978can}. CiM, particularly utilizing memristors technology, offers an innovative alternative. It integrates memory storage and data processing capabilities within a single device, enabling improvements in energy efficiency, speed, and scalability by minimizing data movement and enabling analog computation~\cite{zidan2018future, Prezioso2015, ielmini2018memory}.

In this work, we aim to identify which neural network architectures are scalable, accurate, and hardware-compatible for QST, particularly as quantum systems grow in size and complexity. To that end, we comprehensively benchmarked a diverse set of neural network architectures, supervised models (CNN, FCN, RNN, CGAN, Transformer) and unsupervised models (RBM, SVAE), to assess their suitability for reconstructing high-dimensional quantum states. Beyond simply comparing performance, our goal was to uncover which models maintain high fidelity, converge quickly, and scale favorably as the number of qubits and measurement complexity increase. Among them, CGAN and CNNs consistently outperform others, achieving fidelity up to 0.995 while offering fast convergence and computational efficiency. We report, additionally, the application of the Spiking Variational Autoencoder (SVAE) to QST. Unlike previous DNN-based models, SVAE leverages a sparse, event-driven architecture inspired by neuromorphic computing. Our results show that SVAE achieves high reconstruction fidelity while requiring significantly fewer computational resources. This makes it a strong candidate for future QST platforms, such as edge or embedded quantum diagnostic tools.

\pagestyle{plain}

\begin{figure*}[t!] 
\includegraphics[width=1\linewidth]{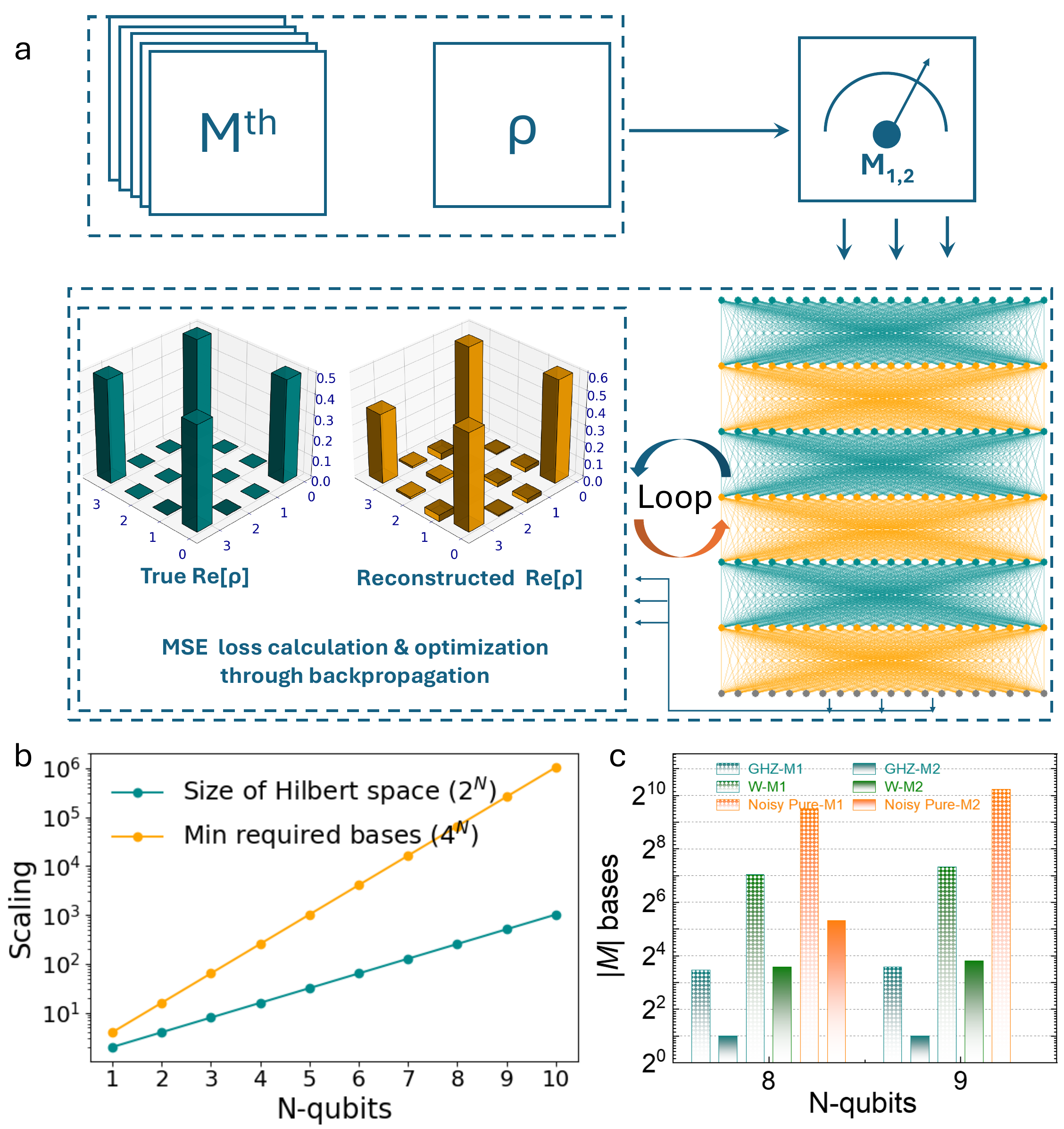}
   \caption{\textbf{Overview of Neural Network-based QST.}  
    \textbf{(a)} Neural Network-based workflow for reconstructing quantum states, optimizing through MSE loss calculation and backpropagation.
    \textbf{(b)} (b) Scaling of Hilbert space size ($2^N$) and the full Pauli basis size ($4^N$) with increasing qubit number $N$. Note that $4^N$ corresponds to the size of the complete Pauli measurement basis used in standard informationally complete tomography, and does not represent the practical minimal number of measurement settings required for all states.
    \textbf{(c)} Measurement bases required for high-fidelity ($\approx 0.99$) reconstruction of GHZ, W, and noisy pure states using expectation-based (M1) and probability-based (M2) measurement methods. Exact values are summarized in Table~\ref{tab:measurement_bases}.}
    \label{fig:fig1}
\end{figure*}

\section*{Background}
\subsection*{Measurement Formalism}
\label{sec:measurement_formalism}

In QST, measurement data are obtained by performing well-defined quantum measurements on an ensemble of identically prepared quantum states. Among the most widely used schemes, particularly in theoretical QST, are \textit{projective measurements}, which correspond to projections of a quantum state onto the eigenbasis of a Hermitian operator.
For single-qubit systems, the standard measurement basis is defined by the Pauli operator set:
\begin{equation}
    \{\sigma_x, \sigma_y, \sigma_z, \mathbb{I}_2\},
\end{equation}
where \(\mathbb{I}_2\) is the \(2\times2\) identity operator. These operators form a complete orthonormal basis for the space of Hermitian operators acting on \(\mathbb{C}^2\).
For \(N\)-qubit systems, the Pauli basis generalizes via tensor products of single-qubit operators, yielding \(4^N\) distinct measurement operators. For instance, for two qubits, a representative element is:
\begin{equation}
    \sigma_z \otimes \sigma_z.
\end{equation}

Projective measurements are a special case of a more general framework known as \textit{Positive Operator-Valued Measures} (POVMs). A POVM is defined by a collection of positive semi-definite operators \(\{M_a\}\) satisfying the completeness relation:
\begin{equation}
    \sum_a M_a = \mathbb{I}_d,
    \label{eq:povm_sum}
\end{equation}
where \(\mathbb{I}_d\) is the identity operator on the Hilbert space of dimension \(d = 2^N\). The probability of obtaining outcome \(a\) when measuring the quantum state \(\rho\) is given by:
\begin{equation}
    P(a) = \mathrm{Tr}(\rho M_a).
    \label{eq:povm_prob}
\end{equation}

To enable full quantum state reconstruction, the measurement operators must be \textit{informationally complete} (\textit{IC}), meaning that their statistical outcomes are sufficient to uniquely determine any \(\rho\). A set of measurement operators \(\{M_a\}\) is \textit{IC} if it spans the space of linear operators on \(\mathcal{H}_d\). That is, any operator \(|\lambda\rangle\) in this space can be expressed as a linear combination of the measurement vectors:
\begin{equation}
    |\lambda\rangle = a |\alpha\rangle + b |\beta\rangle + c |\gamma\rangle + \cdots.
    \label{eq:span}
\end{equation}
In practice, due to finite sampling and noise, this reconstruction is achieved only approximately, and the accuracy depends on the number of measurements, qubit decoherence, and the reconstruction method used.
In QST experiments, a large number of identically prepared quantum systems are measured under different settings to gather statistics. The two main types of information extracted from such measurements are:
i) The \textit{expectation value} of observables.
ii) The \textit{probability distribution} over measurement outcomes.
For a pure quantum state \(|\psi\rangle\), the expectation value of a Hermitian observable \(\hat{A}\) is given by:
\begin{equation}
    \langle A \rangle = \langle \psi | \hat{A} | \psi \rangle.
    \label{eq:expectation_pure}
\end{equation}
For a mixed state described by a density matrix \(\rho\), this generalizes to:
\begin{equation}
    \langle A \rangle = \mathrm{Tr}(\rho \hat{A}).
    \label{eq:expectation_mixed}
\end{equation}

These expressions return the average eigenvalue associated with the measurement of \(\hat{A}\). For instance, computing \(\langle \sigma_x \otimes \sigma_x \rangle\) reveals the expectation value for a two-qubit measurement in the \(X\)-basis.

In parallel, one can analyze the full probability distribution of outcomes. For a pure state \(|\psi\rangle\), the probability of finding the system in eigenstate \(|a\rangle\) is:
\begin{equation}
    P(|a\rangle) = |\langle a | \psi \rangle|^2.
    \label{eq:probability_pure}
\end{equation}
For mixed states, this probability becomes:
\begin{equation}
    P(|a\rangle) = \mathrm{Tr}(\rho |a\rangle \langle a|).
    \label{eq:probability_mixed}
\end{equation}

These measurement statistics, expectation values or full probabilities, form the foundation of quantum state reconstruction. Whether via maximum likelihood estimation, Bayesian inference, or machine-learning–based techniques, all QST methods ultimately rely on the informational completeness of the chosen measurement protocol.

\begin{figure*}[t!]
    \centering
    \includegraphics[width=1.0\linewidth]{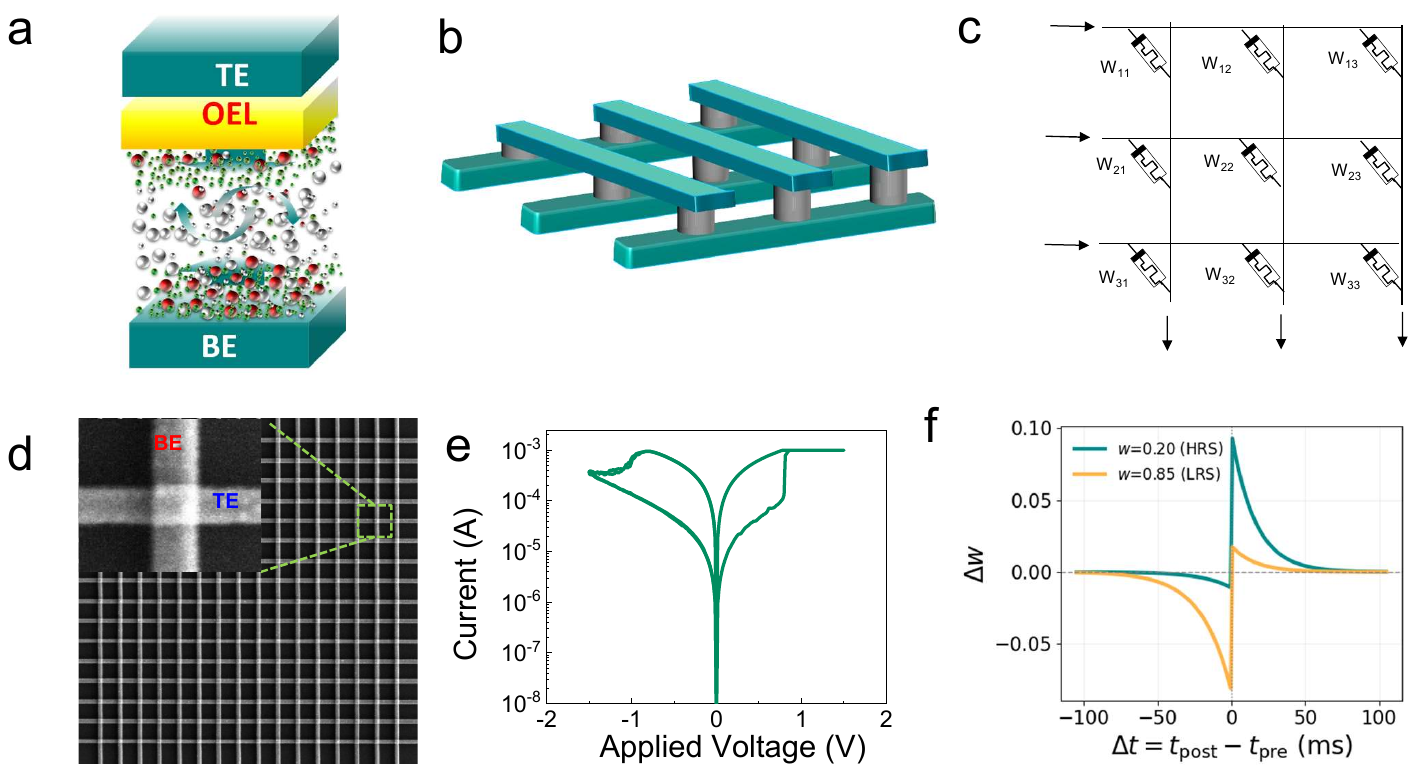}
    \caption{\textbf{(a)} Illustrative diagram of a memristor. TE: top electrodes; OEL: oxygen change layer; BE: bottom electrodes;  the red, green and grey dots represent metals from BE or TE, oxygen ions and oxygen vacancies, respectively. \textbf{(b)} Schematic crossbar array of memristor 3 $\times$ 3 and its circuitry representation for MVM computation (\textbf{(c)}). \textbf{(d)} Scanning electron Microscopy (SEM) images of a real fabricated 16 $\times$ 16 memristor crossbar array. \textbf{(e)} I-V plot of a memristor. \textbf{(f)} STDP learning rules for synaptic plasticity}
    \label{Figure_2}
\end{figure*}

\subsection*{Neural Network Architectures and Learning Paradigms}

To address the challenge of reconstructing quantum states from measurement data, we explore artificial neural networks as efficient learning-based models.  We consider seven representative neural networks architectures in this study: CNNs, FCNs, RNNs, RBMs, CGANs, Transformers and SVAE, chosen to span a broad spectrum of learning paradigms (supervised vs. unsupervised), structural designs (feedforward, recurrent, generative, spiking), and application strengths (e.g., spatial encoding, temporal modeling, distribution learning). For detailed architectural descriptions, see Appendix ~\ref{secA_B}. Each model offers unique inductive biases tailored to specific learning tasks, for instance, CNNs for spatially structured inputs, RNNs for sequential data, and Transformers for attention-driven context modeling.

While architectural design is important, the ultimate performance of a neural network is predominantly determined by the nature and quality of the training data. Equally crucial is the learning paradigm, such as supervised, unsupervised, or generative training, which is typically intrinsic to the architecture itself and significantly shapes its behavior.
\textit{Supervised Learning:} The most common paradigm, it utilizes labeled datasets to train a model by minimizing a predefined loss function that quantifies the difference between the predicted and actual outputs.
\textit{Unsupervised Learning:} This approach relies on unlabeled data, with the objective of discovering latent structure or statistical patterns, such as correlations, clusters, or low-dimensional manifolds, that characterize the data distribution~\cite{Krenn2023}.

In this work, we focus on supervised and unsupervised learning paradigms for quantum state reconstruction, as they are the most established and practically applicable frameworks in this domain. Reinforcement learning and other paradigms remain less explored in QST and are therefore beyond the scope of this study. In the unsupervised setting, models learn a probability distribution from measurement data and subsequently reconstruct the corresponding quantum state. On contrast, supervised learning allows direct mappings from measurement data to target quantum states, which enables task-specific training objectives and more data-efficient optimization by minimizing supervised loss functions. Among the architectures considered, all neural networks models employ supervised learning, with the exception of the \textit{RBM} and the \textit{SVAE} models, which are trained using unsupervised techniques as listed in the table~\ref{tab:nn-paradigms}.

\begin{table}[ht]
\centering
\caption{Overview of neural network architectures, learning paradigms, and DNN classification.}
\label{tab:nn-paradigms}
\begin{tabular}{lccr}
\hline
\textbf{Model} & \textbf{Learning Paradigm} & \textbf{DNN (Y/N)} & \textbf{Notes} \\
\hline
CNN & Supervised & Yes & Spatial inductive bias \\
FCN & Supervised & Yes & Fully connected layers \\
RNN & Supervised & Yes & Temporal sequences \\
CGAN & Supervised & Yes & Generative supervised mapping \\
Transformer & Supervised & Yes & Attention mechanism \\
RBM & Unsupervised & No & Generative, probabilistic \\
SVAE & Unsupervised & No & Spiking, event-driven encoding \\
\hline
\end{tabular}
\end{table}

Training of neural networks proceeds through two fundamental phases: \textit{feed-forward computation} and \textit{backpropagation}. In the feed-forward phase, input data are propagated through the network layers to generate an output. In the backpropagation phase, the model prediction is compared against the ground truth using a loss function, and the resulting error gradient is propagated backward through the network to update the model parameters. This process is repeated iteratively until convergence, i.e., when the loss is minimized below a specified threshold \cite{Yi2023}. The workflow is illustrated in Fig.~\ref{fig:fig1}(a), where it provides  neural network–based approach to QST. To operationalize this, we outline the neural network training process for QST in Fig~\ref{fig:fig1}(a). Firstly, theoretical measurement bases (\( M^{th} \)) and density matrix (\( \rho \)) are used to generate simulated measurement data (\( M_{1,2} \)). This data is fed into the neural network, which outputs the reconstructed state (\( \rho \))  as shown in 3D plots which illustrate true and reconstructed real parts (\(\text{Re}[\rho]\)) of  density matrices for fidelity evaluation.
Optimization of the network parameters is typically performed using stochastic gradient descent (\textit{SGD}). In practice, we employ the \textit{Adam optimizer}, a variant of \textit{SGD} that uses first-order gradient estimates combined with adaptive learning rates and moment estimates for more efficient convergence. Adam requires relatively little memory and is widely used for deep learning applications \cite{Kingma2014}.

The choice of loss function is application-specific. For quantum state reconstruction, we adopt the commonly used \textit{Mean Squared Error (MSE)} loss, which computes the average of the squared differences between predicted outputs \(\hat{y}_i\) and true values \(y_i\) across a dataset of \(N\) samples:
\begin{equation}
    \mathrm{MSE} = \frac{1}{N} \sum_{i=1}^{N} (y_i - \hat{y}_i)^2.
    \label{MSE_loss}
\end{equation}

\begin{equation}
\mathbb{E}_{y \sim p_{\text{data}}} \left[ \log D(y; \theta_D) \right] 
+ \mathbb{E}_{z \sim p_z} \left[ \log \left( 1 - D(G(z; \theta_G); \theta_D) \right) \right]
\label{eq:cgan_loss}
\end{equation}

Those loss functions are both simple to implement and analytically tractable, making it a natural choice for regression-type learning problems such as quantum state estimation. The methodology used for reconstructing quantum states in this study involves training a neural network to map measurement data to a target quantum state. In the pre-processing stage, for example, a 3-qubit GHZ state from equation~\ref{GHZ_state} and Pauli basis $XYZ$, $XIX$ and $ZXY$ are generated. Subsequently, measurements are performed on the GHZ state with the operators using eqation~\ref{eq:expectation_pure},\ref{eq:expectation_mixed} to compute the expectation value of measuring the three observables and equation~\ref{eq:probability_pure}, \ref{eq:probability_mixed} to compute the probability of finding a quantum state in the eigenstate $|a \rangle$. In the training loop stage the resulting measurement data is used as input to a neural network, which after transformations performed by the hidden layers will output the complex coefficients of the reconstructed quantum state. Measurements are performed on the reconstructed state and used to compare to the true measurement data obtained during the pre-processing stage. Subsequently, the measurement outcomes are used to minimize the MSE loss function from equation~\ref{MSE_loss} or equation~\ref{eq:cgan_loss} ( tailored for CGAN~\cite{ahmed2021quantum}) during each loop of the training stage.

\begin{figure*}[t!]
\centering
\includegraphics[width=0.95\linewidth]{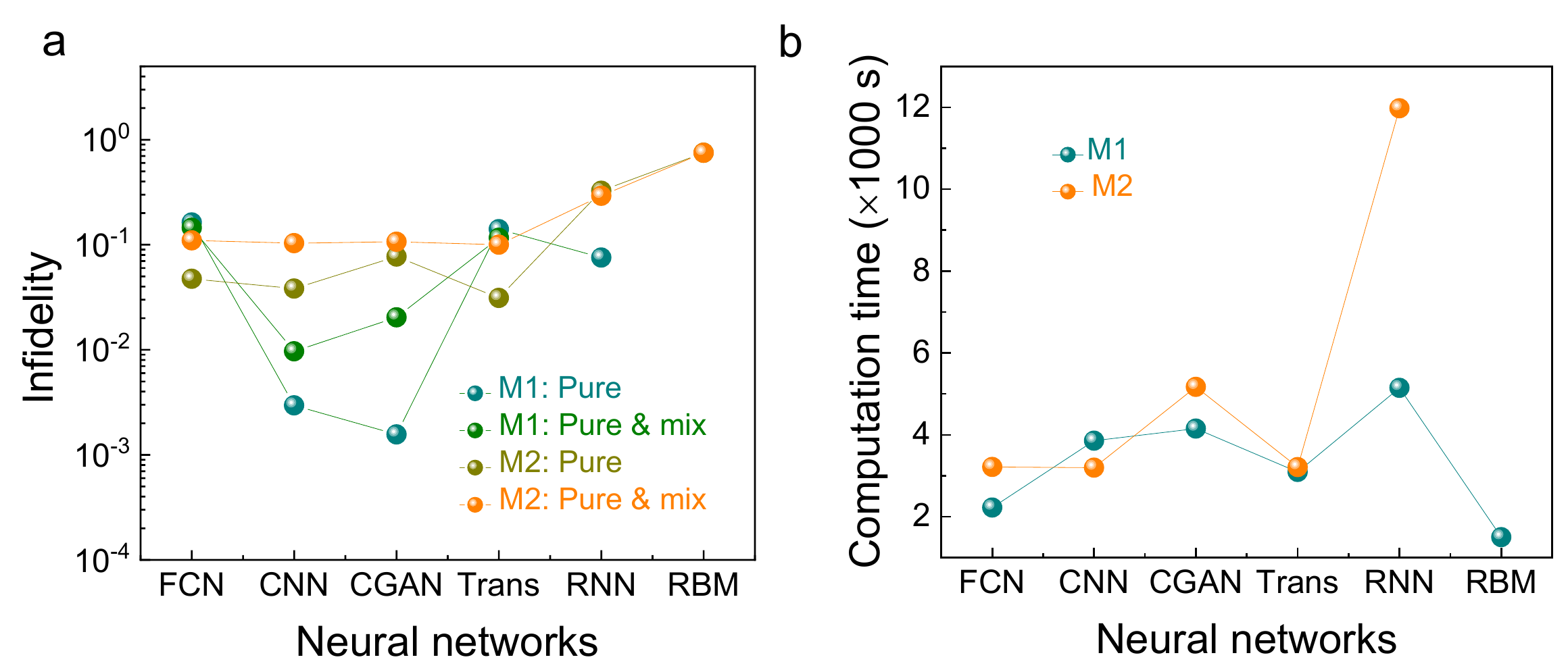}
\caption{The average reconstruction \text{infidelity} \textbf{(a)} and corresponding computation time \textbf{(b)} across various neural network models, including FCN,CNN, CGAN, Trans, RNN, and RBM, for both pure and mixed quantum states under the two measurement methods with 8 qubits over 100 iterations.}
\label{Figure_3}
\end{figure*}

\subsection*{Memristor-based Energy-efficient Computing for Scalable QST}

Computation-in-Memory (CiM) is a promising paradigm designed to overcome the memory wall problem associated with conventional von Neumann architectures. Traditional systems require frequent data transfers between memory and processor, resulting in significant latency and energy inefficiency. In contrast, CiM architectures enable both storage and computation in the same physical location, thereby reducing data movement and improving computational throughput.

Memristor is a two-terminal resistive device that naturally aligns with the CiM paradigm. These devices can function as both memory and computational units, making them ideal candidates for energy-efficient, non-von Neumann architectures. Compared to CMOS technology, it offers key advantages including non-volatility, low power consumption, small footprint, high scalability, and fast analog computation capabilities~\cite{Cai2019,Dash2021}. Figure~\ref{Figure_2}(a) shows the basic structure of a memristor, comprising a metal/insulator/metal stack where the insulating layer (typically an oxide) is sandwiched between top (TE) and bottom (BE) electrodes. Figures~\ref{Figure_2}(b) and (c) illustrate a 3$\times$3 memristor crossbar array designed to perform matrix-vector multiplication (MVM), a core operation in neural networks. Mathematically, MVM is given by:

\begin{equation}
    \mathbf{y} = \mathbf{W} \cdot \mathbf{x}
    \label{eq:mvm_eq}
\end{equation}

In this architecture: (i) Each memory cell stores a weight ($\mathbf{W}$) as its conductance; (ii) The input vector $\mathbf{x}$ is applied as voltages across word lines. (iii) The resulting current at each bit line inherently performs analog multiply-and-accumulate (MAC) operations governed by Ohm's and Kirchhoff’s laws. Figure~\ref{Figure_2}(d) shows an SEM image of a fabricated 16$\times$16 crossbar array with 100 nm $\times$ 100 nm node dimension, demonstrating physical feasibility. Compared to digital hardware, such analog computation offers key advantages for VMM-intensive applications like QST. Specifically, memristor-based CiM that is used to perform VMMs has the following potential:
(i) \textit{Reduced computation time:} Analog MAC operations replace sequential digital steps, allowing parallel execution of entire matrix-vector operations in a single cycle.
(ii) \textit{Lower energy consumption:} By eliminating the need for memory access and reducing data movement, memristor-based VMM consumes significantly less power per operation.
(iii) \textit{Massive parallelism:} All weights and inputs are operated on simultaneously in the crossbar, ideal for the parallel nature of quantum state reconstructions.
(iv) \textit{Improved scalability:} As the number of qubits increases, so does the model complexity. memristor small footprint and stackable architecture allow scaling to meet these growing demands.

In QST, neural networks are trained to reconstruct high-dimensional quantum states from measurement data. As qubit number $N$ increases, both the Hilbert space ($2^N$) and required measurement bases ($4^N$) grow exponentially. The efficiency of analog MVMs using CiM hardware thus becomes essential to sustain this scalability. memristor devices also exhibit binary resistance states: a low-resistance state (LRS, logic \texttt{1}) and a high-resistance state (HRS, logic \texttt{0}), as shown in Figure~\ref{Figure_2}(e). Transitions between these states via SET and RESET operations underpin their functionality for storage and computation. Beyond inference, neuromorphic computing using memristor supports on-chip learning through spike-timing-dependent plasticity (STDP)~\cite{Cai2019,Dash2021}. As illustrated in Figure~\ref{Figure_2}(f), memristor synapses adjust conductance based on temporal patterns of neural activity. Applied to QST, this supports:
(i) \textit{Real-time adaptive learning:} STDP enables QST networks to be updated on-chip as new quantum measurements are obtained.
(ii) \textit{Energy-efficient optimization:} Local weight adaptation removes the need for high-latency, high-power global updates.
(iii) \textit{Scalable deployment:} Embedded STDP learning within memristor makes it feasible to deploy self-improving QST systems as the quantum system scales.
Thus, by leveraging both analog inference and local learning, memristor-based CiM and STDP mechanisms align tightly with the computational demands of QST. This synergy offers a robust and energy-efficient hardware substrate for building scalable QST engines.

\section*{Results}

\subsection*{Measurement bases}

The actual number of measurement bases $|M|$ needed for accurate QST depends not only on the choice of measurement strategy, such as M1 or M2, but also on the type of quantum state, the neural network architecture used for reconstruction, and the target fidelity (e.g., up to 99\%). Since the optimal $|M|$ for a given NN and state is generally unknown, this section empirically evaluates how these factors influence measurement requirements.

To understand the challenge of scalability, Fig~\ref{fig:fig1}(b) visualizes the scaling of the Hilbert space dimensionality (\(2^N\)) and the minimal required measurement bases (\(4^N\)) as functions of qubit number \(N\). This \(4^N\) bound represents the worst case for arbitrary mixed states, whereas compressed-sensing methods can achieve informational completeness with far fewer measurements for low-rank or nearly pure states~\cite{gross2010quantum,flammia2012quantum,kalev2015quantum}. In this work, we restrict our analysis to standard Pauli-basis tomography, where the $4^N$ bound provides a useful reference, but we emphasize that alternative approaches can achieve informational completeness more efficiently when prior structure is exploited. We also compare methods M1 and M2 to understand how the type of measurement data affects the number of required measurement bases $|M|$ for accurate QST. Fig~\ref{fig:fig1}(c) compares measurement bases required to achieve high fidelity (\(\approx 0.99\)) reconstruction forGHZ states, W-states, and noisy pure states, showing that M2 (probability distributions) requires substantially fewer bases. We use two types of methods to acquire measurement data because they reflect common practical approaches in QST research and offer a tradeoff between computational complexity and reconstruction performance: \textit{M1:}  Compute true expectation values $\hat A$ for the set of measurement bases $|M|$ using equation \ref{eq:expectation_pure} for pure states $| \psi \rangle$ and equation \ref{eq:expectation_mixed} for mixed states $\rho$.
\textit{M2:} Compute true probabilities for measuring eigenstates $| a_i \rangle$ using the $M$ sets of measurement bases with equation \ref{eq:probability_pure} for pure states $| \psi \rangle$ and equation \ref{eq:probability_mixed} for mixed states $\rho$.

Using measurement bases that result in an expectation value of zero creates instability in the reconstruction process. This instability arises because the measured outcomes fluctuate equally between $+1$ and $-1$, which reduces the signal-to-noise ratio and makes the neural network estimation of the expectation value highly sensitive to small  sampling fluctuations. Our numerical experiments (see Fig.~\ref{fig:fig1}(c)) show that including these zero-expectation bases does not significantly improve the reconstruction fidelity for method M1. In fact, M1 is particularly affected because it relies on a single scalar expectation value per measurement basis. Measurement bases yielding expectation values near zero add little information and increase noise sensitivity, particularly for M1. Therefore, we restrict our analysis to bases with non-zero expectation values. However, as shown in Fig.~\ref{fig:fig1}(b), QST is fundamentally limited by the exponential scaling of the Hilbert space ($2^N$) and the minimum required number of measurement bases ($4^N$) to be informationally complete~\cite{Carrasquilla2018}. Fig.~\ref{fig:fig1}(c) then evaluates how many measurement bases with non-zero expectation values are required to fully reconstruct three representative pure quantum states using methods M1 and M2. This allows us to explicitly connect the informational content of the measurement bases with the empirical reconstruction requirements.

To place the measurement requirements into context, we now evaluate how the  number of measurement bases $|M|$ scales for representative quantum states. Our goal is to assess whether sub-exponential scaling in $|M|$ and the associated measurement data can be achieved in practice, compared to the exponential upper bound of $4^N$ required for informational completeness standard Pauli-basis tomography. For $N = 8$ qubits, the  maximum number of measurement bases in the Pauli basis is $4^8 = 65,536$, and  for $N = 9$ it is $4^9 = 262,144$. Figure~\ref{fig:fig1}(c) summarizes the empirically determined values of $|M|$ required to reach a reconstruction fidelity of approximately $0.99$ for three representative quantum states: a GHZ state, a W state, and a noisy pure state  (see Appendix~\ref{appen:quantum_states} for definitions) and a summary of the empirically required measurement bases $|M|$ for GHZ, W, and noisy pure states using both methods is provided in Table~\ref{tab:measurement_bases}. These $|M|$ values are obtained from our numerical experiments. (reconstruction for $N = 9$ was not feasible within available memory).
We observe that as the number of computational basis states with non-zero amplitudes in the quantum state increases, the number of measurement bases required to fully reconstruct both the amplitudes and phases also increases. Interestingly, the degree of entanglement itself is not the dominant factor:  although the GHZ state is maximally entangled, it is relatively easy to reconstruct, while the W state, despite being less entangled, requires  significantly more measurement bases due to the larger number of  computational basis states with non-zero amplitudes.

\begin{table}[h!]
\centering
\caption{Number of measurement bases $|M|$ required to achieve high-fidelity ($\approx 0.99$) reconstruction for different quantum states using M1 and M2 methods.}
\begin{tabular}{lcccr}
\hline
\multirow{2}{*}{Quantum state} & \multicolumn{2}{c}{M1: Expectation-based} & \multicolumn{2}{c}{M2: Probability-based} \\
 & $N = 8$ & $N = 9$ & $N = 8$ & $N = 9$ \\
\hline
GHZ        & 11   & 12    & 2   & 2   \\
W          & 130  & 160   & 12  & 14  \\
Noisy pure & 720  & 1200  & 40  & --- \\
\hline
\end{tabular}
\label{tab:measurement_bases}
\end{table}

To conclude, M2 requires fewer bases but is experimentally more demanding since full probability distributions need more repeated measurements; M1 converges with fewer samples but requires more distinct bases. As a result, each data set from M2 constrains the possible quantum states more strongly, which in turn reduces the number of distinct measurement bases needed for a given target fidelity. However, while this advantage is clear in simulation (where ideal probability distributions can be directly computed), it is less practical experimentally.  In real experiments, obtaining the full probability distribution for M2 requires significantly more repeated measurements per basis to collect sufficient statistics for each possible outcome. In contrast, M1 only requires repeated measurements to estimate the average expectation value, which typically converges with fewer samples. This makes M1 more scalable in experimental settings, even though it requires more distinct measurement bases to reach the same reconstruction fidelity.

\subsection*{Neural Network Performance Evaluation}

To quantitatively assess the effectiveness of different neural network architectures for QST, we evaluate their performance across two critical metrics: reconstruction accuracy (fidelity) and computational efficiency. To identify which architectures maintain high fidelity while remaining computationally practical as quantum systems scale, and how different measurement strategies, M1 and M2, affect reconstruction outcomes.

As depicted in Fig.~\ref{Figure_3}(a), two of six architectures CNN and CGAN consistently achieve the minimal infidelity values, with CGAN yielding the highest fidelity across all settings. Specifically, CGAN and CNN with M1 on pure states achieves the best reconstruction performance with a infidelity lower than $2 \times 10^{-3}$. In contrast, RBM and RNN perform poorly, especially for mixed states using M2, with infidelity values exceeding $10^{-1}$. While M2 is theoretically richer in information content, it leads to slightly higher reconstruction infidelity across most architectures, likely due to increased input complexity and learning instability when processing full probability distributions.
In terms of computation Time as shown in Fig.~\ref{Figure_3}(b), FCN and CNN demonstrate the shortest training durations (around 2,000 seconds), making them ideal for practical applications. CGAN and Transformer exhibit moderate computational demands (3,000–6,000 seconds), while RNN and RBM are significantly slower, particularly RBM under M2, which exceeds 12,000 seconds. This reinforces the need to consider both accuracy and efficiency when selecting architectures for real-world QST deployments.
Taking above considerations, these results demonstrate that CNN offers the most balanced performance in terms of fidelity and speed. Although M2 theoretically provides more informative measurement data, M1 results in more stable and efficient training across most models, especially for mixed quantum states. This insight is crucial for guiding experimental and hardware-constrained implementations of neural-network-based QST.

\begin{figure*}[t!]
\centering
\includegraphics[width=0.9\linewidth]{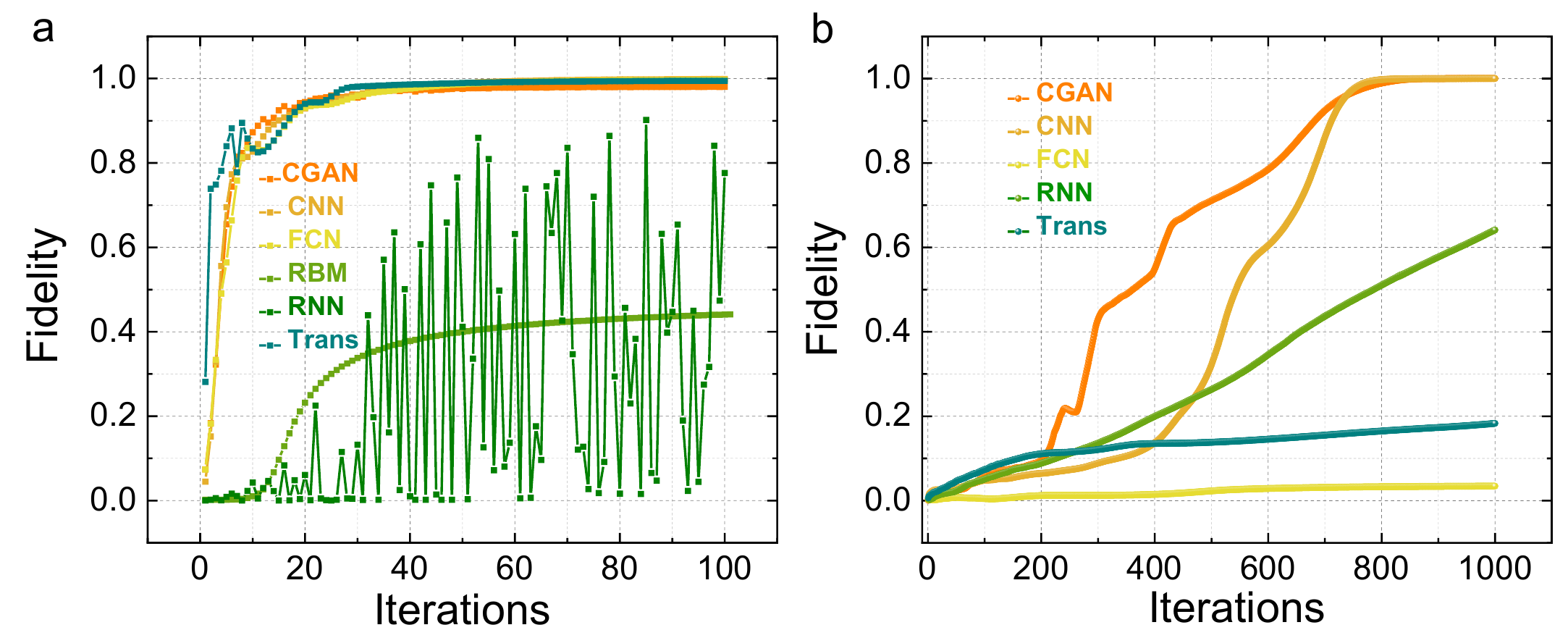}
\caption{\quad Fidelity as a function of iterations for a noisy mixed state. \textbf{(a)} Reconstruction of a pure 9 qubit GHZ state for 6 different neural network architectures with M2, using $2$ measurement bases with non-zero expectation values. \textbf{(b)} Reconstruction process of a pure $9$ qubit noisy state for $5$ different neural network architectures with M1, using $1,200$ measurement bases with non-zero expectation values.}
\label{Figure_4}
\end{figure*}

\begin{table}
\centering
\caption{Computation time comparison (in seconds) for different neural network architectures using methods M1 and M2.}
\small
\begin{tabular}{lccc}
\toprule
\textbf{Neural Network} & \textbf{M2 (2 bases)} & \textbf{M1 (1200 bases)} \\
\midrule
CGAN   & 710    & 19,468 \\
CNN    & 290    & 19,687 \\
FCN    & 280    & 11,043 \\
RBM    & 15,768 & N.A.    \\
RNN    & 843    & 21,252 \\
Trans  & 303    & 16,798 \\
\bottomrule
\end{tabular}
\label{tab:computation_time}
\end{table}

To illustrate the details of reconstruction process for various neural networks. Figure \ref{Figure_4}(a) shows a 9 qubit GHZ state for different neural network architectures using M2, performed with 2 measurement bases with non-zero expectation values. All neural networks rapidly converge to a fully reconstructed except for RNN and RBM. RNN shows major instability in the large oscillations between fidelity values and does not converge to a high fidelity ($>$ 0.99). RBM only converges to a fidelity of around 0.43. For this case, supervised learning is able to perform significantly better than the unsupervised RBM model. However, while the supervised learning models are Deep Neural Networks (DNNs) with multiple (3 or more) hidden layers, the RBM model is only a single layer. This requires more research to reliably compare supervised and unsupervised learning models for QST. Among the neural networks, FCN has the  smallest amount of iterations and time for high-fidelity convergence, with CNN and Transformer showing very similar performance as listed in the caption. CGAN requires 2.5 times longer, RNN takes 3 times as long, and RBM is the most time consuming, taking 5.6 times.
In Figure~\ref{Figure_4}(b), a 9-qubit state initially prepared as a pure state is reconstructed after being subjected to noise, resulting in a mixed state. The reconstruction is performed with $|M|=1,200$ measurement bases using M1. Among the tested models, only CNN and CGAN are able to fully reconstruct the mixed state (fidelity $>0.99$) within roughly the same number of iterations. In contrast, RNN, Transformer, and FCN do not reach comparable fidelities within 1000 iterations; FCN, in particular, almost fails to extract meaningful information from the measurement data. Notice how many more iterations are needed here compared to the noiseless case in Figure~\ref{Figure_4}(a); this is due to the increased complexity of the noisy mixed state, which requires more measurement bases and longer training to reconstruct. CNN and CGAN have approximately equal computation times, as listed in Table~\ref{tab:computation_time}.
These results highlight CNN and CGAN as the most robust and scalable supervised architectures, while unsupervised models like RBM struggle to achieve comparable fidelity. These insights provide a practical guideline for selecting models when balancing fidelity, computational time, and scalability in QST. It worth noting that CNN is also suitable for CiM architecture for neural network acceleration~\cite{yao2020fully,aguirre2024hardware}, which is also potentially applied in the hardware acceleration for QST.
 
\begin{figure*}[t!]
\centering
\includegraphics[width=0.9\linewidth]{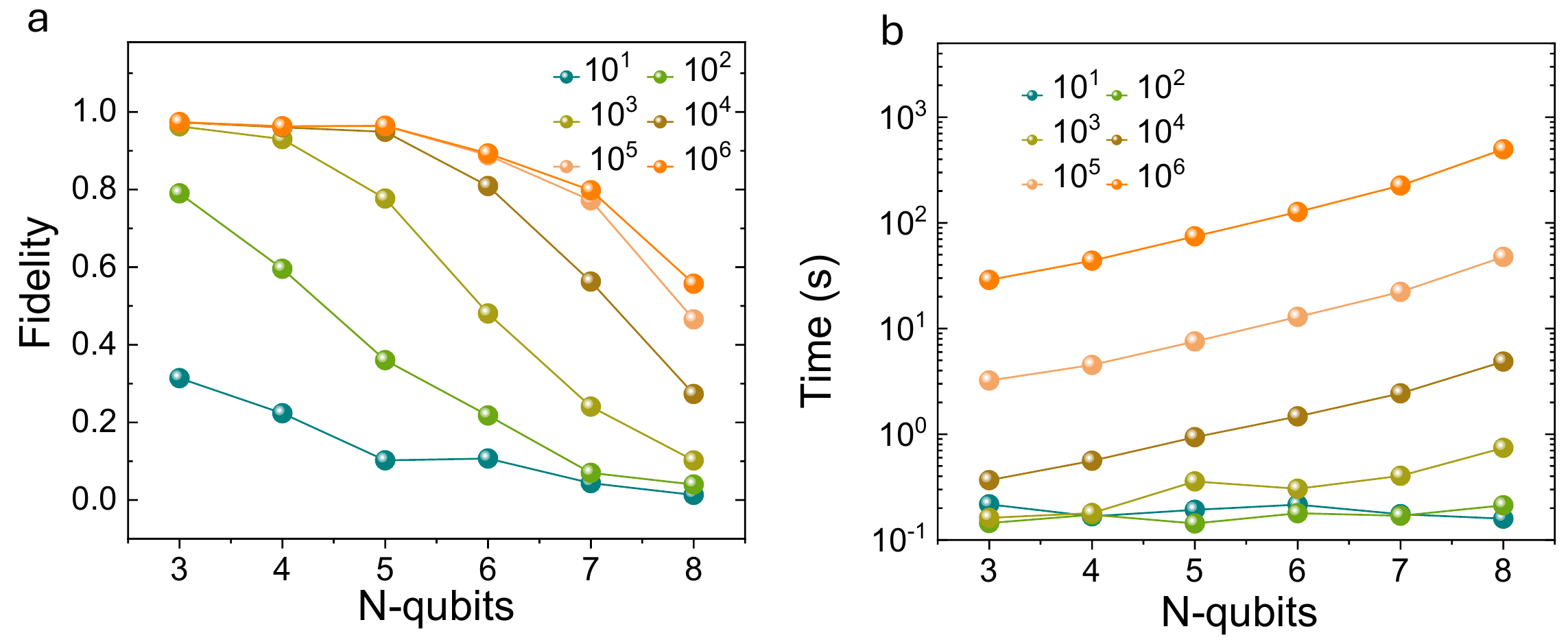}
\caption{\quad\textbf{Performance evaluation of SVAE neural network for QST.} 
\textbf{(a)} Fidelity of the reconstructed quantum states as a function of the number of qubits, evaluated over six different total measurement shots (repetitions/samples): $10^1$, $10^2$, $10^3$, $10^4$, $10^5$, and $10^6$. 
\textbf{(b)} Corresponding computational time required for the neural network training as a function of qubit number, under the same shots.}
\label{qsvae}
\end{figure*}

Because the SVAE uses an event-driven spiking architecture, it can be naturally mapped to neuromorphic hardware~\cite{czischek2022spiking,klassert2022variational}. This makes it more hardware-friendly compared to standard CNNs or Transformers. Meanwhile, SVAE differs fundamentally from deterministic supervised models in both learning paradigm and hardware relevance, and shows unique fidelity-scaling trends, we evaluate it separately to highlight its strengths and limitations for scalable QST.
Figure~\ref{qsvae} presents an in-depth performance analysis of the \textit{SVAE} architecture applied to QST using pure GHZ states with M2. The model performance is evaluated across two principal dimensions: the fidelity of reconstructed quantum states and the computational time required, both as a function of the number of qubits and the total number of measurement counts or shots. Figure~\ref{qsvae}(a) shows the fidelity of the SVAE-generated quantum state reconstructions as a function of qubit number, evaluated for six total shot counts ranging from $10^1$ to $10^6$. 

At low shot counts ($10^1$–$10^2$), the SVAE exhibits poor performance above 4 qubits, with fidelities rapidly decaying to near-zero values. This behavior is attributable to insufficient statistical sampling in high-dimensional Hilbert spaces, where the state space grows exponentially as $2^N$ for an $N$-qubit system. Without enough measurement data, the SVAE lacks the information required to learn a faithful generative distribution, resulting in high reconstruction error.
As the number of measurement shots increases, the model performance improves markedly. For $10^5$ and $10^6$ shots, the SVAE consistently achieves fidelities above 0.9 for systems up to 6 qubits. These results highlight the SVAE capacity to utilize rich statistical information effectively. The saturation behavior observed in fidelity for high shot counts suggests that the model error becomes bounded by its expressivity rather than data limitations. Figure~\ref{qsvae}b displays the computational time required for the SVAE to perform quantum state reconstructions, plotted as a function of qubit number and shot count. For small systems (3 to 5 qubits), the inference time remains low, even at the highest shot levels. This computational efficiency stems from the SVAE ability to perform amortized inference, whereby the decoder maps latent representations to full density matrices without requiring iterative optimization for each individual measurement. 

As both the number of qubits and the total shot count increase, the computational cost grows progressively. Larger datasets necessitate longer data loading and preprocessing times, and deeper networks or higher-dimensional latent spaces require more iterations during training and evaluation. For 7 to 8 qubits with $10^6$ shots, the time cost reaches the hundreds of seconds range, reflecting increased optimization complexity and the growing burden of processing high-dimensional input features.
Unlike deterministic neural networks such as CNN, CGAN, and Transformer, the SVAE leverages its generative latent space to effectively model uncertainty and incomplete measurement data. Its probabilistic framework provides greater robustness under noisy or data-sparse conditions, where deterministic models often struggle. Moreover, the event-driven spiking architecture of SVAE makes it inherently more energy-efficient and hardware-friendly for neuromorphic and memristor-based CiM implementations, offering a scalable pathway for on-chip QST in resource-constrained environments.

Compared to traditional QST methods such as Maximum Likelihood Estimation (MLE) or Bayesian techniques, which scale exponentially with qubit number and become impractical beyond roughly 6--8 qubits~\cite{patel2025selective,ahmad2022self}, the SVAE demonstrates a more favorable scaling profile. In our experiments, SVAE reliably reconstructed pure GHZ states up to 7 qubits using $10^5$--$10^6$ shots, whereas MLE would require a fully informationally complete measurement set scaling as $4^N$. Although fidelity drops at very low shot counts ($10^1$--$10^2$) for systems above 4 qubits (as shown in Fig.~\ref{qsvae}), SVAE generative modeling enables it to manage higher-dimensional states with fewer measurements and less computational overhead~\cite{rocchetto2018learning}. This balance positions SVAE as a promising candidate for near-term quantum experiments that require real-time feedback and practical reconstruction fidelity.

Figure~\ref{qsvae}~a,b illustrate a key trade-off in QST using SVAE models: fidelity improves with larger data availability, but at the cost of higher computation time. At high shot counts ($10^5$--$10^6$), the SVAE achieves the reasonable fidelities observed in our experiments; however, fidelity consistently declines as the number of qubits increases, and does not exceed 0.9 for systems more than 6 qubits. At lower shot counts ($10^1$--$10^2$), the fidelity rapidly decreases for all qubit numbers, reflecting the severe information deficit when measurement data are sparse. These results underline that, while SVAE benefits from richer data, scalability remains a major challenge for systems with many qubits. One distinctive advantage of the SVAE, compared to deterministic neural network models such as CNNs, CGANs, or Transformers, is its spiking, event-driven architecture. This makes the SVAE inherently more hardware-friendly and energy-efficient, as it can be mapped onto neuromorphic platforms such as memristor-based CiM accelerators. Such compatibility positions SVAE as a promising candidate for energy-efficient, on-chip quantum state reconstruction in near-term experiments. These results reinforce broader conclusions in the literature that deep generative models with variational inference represent a compelling direction for mitigating the scalability barriers of conventional QST, while offering opportunities for more hardware-efficient implementations.

\section*{Discussion}

This work presented a systematic benchmarking of neural network (NN) architectures for QST using two distinct measurement methodologies (M1 and M2). The comprehensiveness of the analysis stems from the inclusion of models across the major NN paradigms: FCN, CNN, RNN, transformers, CGAN, RBM, and SVAE. These models span supervised and unsupervised learning, deterministic and generative inference, and architectures with varying scalability and hardware compatibility, ensuring that the comparisons reflect the state of the art in machine learning for QST.
Several key observations emerge from this analysis. CGAN and CNN architectures consistently achieve the best balance of reconstruction fidelity and computational efficiency across pure and mixed states, confirming their suitability for large-scale QST. In contrast, RBM and RNN models are highly sensitive to hyperparameters (e.g., learning rate) and exhibit poor scalability, often failing to converge for larger qubit systems or mixed states. SVAE, a generative unsupervised model, shows distinctive behavior: at high shot counts ($10^5$–$10^6$), it attains competitive fidelities for small systems but fidelity declines steadily with increasing qubit number. Nonetheless, SVAE demonstrates enhanced robustness under noisy and data-sparse conditions, where deterministic supervised models degrade more severely. Furthermore, its spiking, event-driven architecture makes it inherently compatible with neuromorphic and memristor-based CiM (CiM) accelerators, an advantage not shared by CGAN.

The comparison of measurement methodologies revealed that while M2 can, in principle, reduce the number of required measurement bases for pure states by exploiting informational completeness, M1 remains more practical for experimental mixed-state scenarios due to it. These findings highlight the interplay between measurement design and algorithmic scalability.
Beyond the algorithmic comparisons, we propose that future work could explore the mapping of CNN and SVAE architectures onto memristor-based CiM platforms. Such hardware-aware integration could, in principle, alleviate the computational bottlenecks of von Neumann architectures by reducing data movement, leveraging non-volatility, and exploiting analog computation. While this study did not include hardware-level simulations or implementations, the unique architectural properties of SVAE and CNN suggest that they are strong candidates for co-design with CiM accelerators to enable scalable, energy-efficient, \textit{in-situ} QST pipelines.

Taking all above considerations, this study establishes CNN, CGAN as the most robust supervised architectures for QST, SVAE as a promising generative alternative with unique hardware compatibility, and M1 measurement strategies as the most practical for mixed-state reconstructions. By leveraging these insights and pursuing architecture–hardware co-design, it would bridge the gap between algorithmic performance and hardware constraints, enabling scalable and energy-efficient quantum state tomography in near-term quantum experiments. While the present study provides a comprehensive benchmarking of neural network architectures and measurement strategies for quantum state tomography (QST), several limitations remain. 
First, hardware-level simulations of memristor-based Computation-in-Memory (CiM) accelerators have not yet been performed, and thus the energy and latency benefits are estimated rather than empirically validated. 
Second, the SVAE architecture, although hardware-friendly, exhibits fidelity degradation when scaling beyond six qubits, requiring further algorithmic improvements for large-scale applications. 
Finally, compressed-sensing techniques were not combined with the neural network approaches in this work; exploring such hybrid methods could significantly reduce measurement requirements while maintaining high reconstruction fidelity. 
Addressing these limitations will be a priority for future research.

\section*{Methods}

The software used is written in Python version 3.11. The hardware specifications of the used computer are: 16 GB RAM, Intel i5-4460 CPU and GeForce GTX 1660 Super GPU with 6 GB of VRAM. A custom layer in the neural network model is used to extract the reconstructed quantum state in order to get a good measure of how the neural network model is performed by computing the fidelity of the true and reconstructed state. The fidelity, which is a measure used to compute the overlap between two quantum states is commonly used to indicate similarity between the states. The definition we used to calculate the fidelity of a pure state is:
\begin{equation}
    Fidelity = |\langle \psi_1 | \psi_2 \rangle|^2.
\end{equation}

The definition we used to calculate the fidelity of a mixed state is:
\begin{equation}
    Fidelity = \left( \Tr\left[\sqrt{\sqrt{\rho_1}\rho_2\sqrt{\rho_1}}\,\right] \right)^2.
\end{equation}

\subsection*{DATA AVAILABILITY}
The data generated for this research is available at \href{https://gitlab.com/StevenvO5/supervised-nns-qst-code}{QST data}.


\subsection*{CODE AVAILABILITY}
The code developed for this research is available at \href{https://gitlab.com/StevenvO5/supervised-nns-qst-code}{QST code}.






\subsection*{AUTHOR CONTRIBUTIONS}
EH: Conceptualization, results analysis, visualization, manuscript draft and finalization (Lead). SO:  Conceptualization, simulation, results analysis, manuscript draft. KYY: Conceptualization, results analysis, visualization, manuscript draft. JL: Simulation, results analysis. RB: Conceptualization, analysis, manuscript draft, supervision. HA: Conceptualization, analysis, manuscript draft, supervision. SN: Conceptualization, results analysis, supervision. SF: Results analysis, visualization, manuscript draft and finalization. RI: Conceptualization, results analysis, supervision, funding collection. All coauthors contribute to the manuscript review and revision. 

\subsection*{COMPETING INTERESTS}
The authors declare no competing interests.

\subsection*{ADDITIONAL INFORMATION}
\subsection*{Supplementary information} 
The online version contains supplementary material available at
https://doi.org/xxx/xxxx.



\begin{appendices}

\section{Quantum states}
\label{appen:quantum_states}
In general a quantum state consists of complex coefficients which contain two types of information: amplitude and phase information. The physical pure quantum states $\rho$ (with  $\rho^2 = \rho$) of interest are the Greenberger-Horne-Zeilinger (GHZ) state, Wolfgang (W) state and a noisy state. The GHZ state is maximally entangled for all number of $N$ qubits with a purity of $1$. The GHZ state is defined as:
\begin{equation}
    |GHZ \rangle = \frac{|0 \rangle^{\otimes N} + |1 \rangle^{\otimes N}}{\sqrt{2}}.
    \label{GHZ_state}
\end{equation}

The W state contains a superposition of N qubits in which only one qubit in every ket is in the $|1 \rangle$ state and the amount of entanglement decreases with increasing amount of qubits $N$. This state is maximally entangled for 2 qubits, partially entangled for 3 qubits and as the number of qubits $N$ increases the degree of entanglement decreases. The W state is defined as:
\begin{equation}
    |W \rangle = \frac{|100...0 \rangle + |010...0 \rangle + ... + |00...01 \rangle}{\sqrt{N}}.
    \label{W_state}
\end{equation}

The noisy pure state consists of randomly generated complex coefficients for every ket and is defined as:
\begin{equation}
    \ket{\psi} = \sum_{i=0}^{2^N-1} c_i \ket{i}.
\end{equation}

In order for the quantum states to be physical states, the corresponding density matrix requires to be positive semi-definite (PSD), hermitian and have $Tr(\rho) = 1$ (normalization).

The physical mixed states of interest are the generalized Werner state.It is defined as:
\begin{equation}
    \rho = p|GHZ \rangle \langle GHZ| + (1-p)I_N/2^N,
    \label{Werner}
\end{equation}

which is a combination of the outer product of a GHZ state and a maximally mixed state coming from the second term $I_N/2^N$. For $p = 0$ it would purely be a maximally mixed state and for $p = 1$ it would purely be a mixed GHZ state. A maximally mixed state contains no amount of entanglement and gives a lower bound on the purity which decreases with increasing number of qubits as $1/2^N$. The overall purity of the Werner state with $p=0.5$ decreases from $0.438$ for 2 qubits to about $0.261$ for 6 qubits and is partially entangled.

The noisy mixed state also consists of randomly generated complex coefficients for every ket and is defined as:
\begin{equation}
    \rho = \sum_{i} p_i \ket{\psi_i}\bra{\psi_i},
\end{equation}
where the density matrix $\rho$ again requires to be PSD, hermitian and have $Tr(\rho) = 1$. Comparing the results from reconstructing these three pure and three mixed quantum states will show the influence of the amount of computational basis states with non-zero amplitudes and the degree of entanglement on the performance of a neural network.

\section{Neural Networks Architectures}\label{secA_B}

\subsection{Fully Connected  Network Model}

\begin{table}[h!]
\centering
\caption{FCN model summary}
\begin{tabular}{lcr}
\toprule
\textbf{Layer type} & \textbf{Output shape} & \textbf{Parameters} \\
\midrule
InputLayer & (None, 4096) & 0 \\
InputLayer & (None, 64, 64, 8192) & 0 \\
Dense & (None, 2048) & 8,388,608 \\
Dense & (None, 2048) & 4,196,352 \\
Dense & (None, 4096) & 8,392,704 \\
Dense & (None, 4096) & 16,781,312 \\
Dense & (None, 8192) & 33,562,624 \\
Reshape & (None, 64, 64, 2) & 0 \\
DensityMatrix & (None, 64, 64) & 0 \\
Expectation & (None, 4096) & 0 \\
\bottomrule
\label{tab:fcn_model}
\end{tabular}
\end{table}

Table \ref{tab:fcn_model} shows the specific layers used in the FCN model and the corresponding output shapes and trainable parameters in the case of $N=6$ qubits and $|M|=4^6 = 4,096$ measurement bases. The first input layer contains the true expectation values for the $|M|=4,096$ sets of measurement bases. The second input layer contains the measurement operators which are only used in the last custom layer Expectation. The shape value $8,192=4,096 \cdot 2$ is due to the separation of real and imaginary parts. Then five consecutive Dense layers are used in order to extract features from the input data and the Dense layers increase in complexity and hence will learn increasingly complex patterns. After each Dense layer a LeakyReLu activation is used to introduce non-linearity into the model which allows the model to learn about more complex features. The learned features are then used to make predictions. The last Dense layer is used for transforming the data into the desired output shape of $2 \cdot 64^2=8,192$ which is required for constructing the density matrix and hence cannot be reduced in size. First, the data needs to be reshaped into the proper form such that it can be passed to the custom DensityMatrix layer in which the reconstructed density matrix is computed. Finally, the expectation layer computes the new expectation values of the reconstructed density matrix.

\subsection{Convolutional Neural Network Model}

\begin{table}[h!]
\centering
\caption{CNN model summary}
\begin{tabular}{lcr}
\toprule
\textbf{Layer type} & \textbf{Output shape} & \textbf{Parameters} \\
\midrule
InputLayer & (None, 4096) & 0 \\
InputLayer & (None, 64, 64, 8192) & 0 \\
Dense & (None, 2048) & 8,388,608 \\
LeakyReLU & (None, 2048) & 0 \\
Reshape & (None, 32, 32, 2) & 0 \\
Conv2DTranspose & (None, 64, 64, 64) & 2,048 \\
InstanceNormalization & (None, 64, 64, 64) & 128 \\
LeakyReLU & (None, 64, 64, 64) & 0 \\
Conv2DTranspose & (None, 64, 64, 64) & 65,536 \\
InstanceNormalization & (None, 64, 64, 64) & 128 \\
LeakyReLU & (None, 64, 64, 64) & 0 \\
Conv2DTranspose & (None, 64, 64, 2048) & 32,768 \\
Conv2DTranspose & (None, 64, 64, 2) & 1,024 \\
DensityMatrix & (None, 64, 64) & 0 \\
Expectation & (None, 4096) & 0 \\
\bottomrule
\label{tab:cnn_model}
\end{tabular}
\end{table}
Table~\ref{tab:cnn_model} shows the specific layers used in the CNN model and the corresponding output shapes and trainable parameters for the case of $N=6$ qubits and $|M|=4^6 = 4,096$ measurement bases. First, the data passes through a Dense layer to increase dimensionality, followed by a LeakyReLU activation to introduce non-linearity into the learning process. The data is then reshaped into a 4D format suitable for the following convolutional layers. 
The Conv2DTranspose layers are used for upscaling the dimensionality of the data and extracting data features, instead of the standard downscaling approach used in typical convolutional layers. This upscaling strategy is commonly employed for data generation or reconstruction tasks. The InstanceNormalization layers are applied after each transpose convolution to keep the data stable and normalized, compensating for the multiple up- and down-scaling steps in the architecture. Finally, the custom \texttt{DensityMatrix} and \texttt{Expectation} layers are appended to generate the final quantum state representation and measurement expectations.

\subsection{Recurrent Neural Network Model}

\begin{table}[h!]
\centering
\caption{RNN model summary}
\begin{tabular}{lcr}
\toprule
\textbf{Layer type} & \textbf{Output shape} & \textbf{Parameters} \\
\midrule
InputLayer (inputs) & (None, 4096) & 0 \\
InputLayer (operators) & (None, 64, 64, 8192) & 0 \\
Reshape & (None, 4096, 1) & 0 \\
SimpleRNN & (None, 4096, 50) & 2,600 \\
SimpleRNN & (None, 50) & 5,050 \\
Dense & (None, 8192) & 417,792 \\
Reshape & (None, 64, 64, 2) & 0 \\
DensityMatrix & (None, 64, 64) & 0 \\
Expectation & (None, 4096) & 0 \\
\bottomrule
\label{tab:rnn_model}
\end{tabular}
\end{table}
Table \ref{tab:rnn_model} shows the specific layers used in the RNN model and the corresponding output shapes and trainable parameters in the case of $N=6$ qubits and $|M|=4^6 = 4,096$ measurement bases. The difference between the FCN model and this one is that the Dense layers are replaced by two SimpleRNN layers. The first SimpleRNN layer processes the input sequentially and feeds the output in a sequence to the second SimpleRNN layer. The second layer processes the sequence of data and returns the final output of the two layers. Here, the Tanh activation function is used, which is more commonly applied in RNN models to introduce non-linearity such that the model is again able to learn more complex patterns.

\subsection{Restricted Boltzmann Machines Model}

\begin{table}[h!]
\centering
\caption{RBM model summary}
\begin{tabular}{lr}
\toprule
\textbf{Model detail} & \textbf{Value} \\
\midrule
Number of visible units & 6 \\
Number of hidden units & 13 \\
Number of parameters in $\lambda$ weights & 162 \\
Number of parameters in $\mu$ weights & 162 \\
Total parameters & 324 \\
\bottomrule
\label{tab:rbm_model}
\end{tabular}
\end{table}
This model uses unsupervised learning and is not part of the DNN category because it only has two layers: a visible and a hidden layer. The RBM model is used to learn a probability distribution from measurement data and subsequently uses the probabilities to reconstruct the quantum state it represents. The parameters used in training the RBM model are shown in Table~\ref{tab:rbm_model}. The RBM model generates counts for every configuration, which are subsequently converted into probabilities. 
Unlike the M2 measurement methodology used in other neural network models, where exact eigenstate probabilities are provided without sampling noise, the RBM uses measurement data that inherently includes statistical fluctuations. The $\lambda$ weights are used for the amplitudes and the $\mu$ weights for the phases of the quantum state, and both sets of parameters are updated in every loop by computing gradients. The number of hidden neurons is set at $2N+1$, and a default of 1,000 counts are generated per basis. The initial learning rate is set high at $0.1$ to prevent the model from being stuck at zero fidelity during early training.

\subsection{Conditional Generative Adversarial Networks Model}
CGAN consists of a generator and a discriminator model. Table \ref{tab:generator_model_summary} shows the specific layers used in the Generator model and the corresponding output shapes and trainable parameters in the case of $N=6$ qubits and $|M|=4^6 = 4,096$ measurement bases. This model is equivalent to the CNN model.

\begin{table}[h!]
\centering
\caption{CGAN: Generator model summary}
\begin{tabular}{lcr}
\toprule
\textbf{Layer type} & \textbf{Output shape} & \textbf{Parameters} \\
\midrule
InputLayer & (None, 4096) & 0 \\
InputLayer & (None, 64, 64, 8192) & 0 \\
Dense & (None, 2048) & 8,388,608 \\
LeakyReLU & (None, 2048) & 0 \\
Reshape & (None, 32, 32, 2) & 0 \\
Conv2DTranspose & (None, 64, 64, 64) & 2,048 \\
InstanceNormalization & (None, 64, 64, 64) & 128 \\
LeakyReLU & (None, 64, 64, 64) & 0 \\
Conv2DTranspose & (None, 64, 64, 64) & 65,536 \\
InstanceNormalization & (None, 64, 64, 64) & 128 \\
LeakyReLU & (None, 64, 64, 64) & 0 \\
Conv2DTranspose & (None, 64, 64, 32) & 32,768 \\
Conv2DTranspose & (None, 64, 64, 2) & 1,024 \\
DensityMatrix & (None, 64, 64) & 0 \\
Expectation & (None, 4096) & 0 \\
\bottomrule
\label{tab:generator_model_summary}
\end{tabular}
\end{table}

\begin{table}[h!]
\centering
\caption{CGAN: Discriminator model summary}
\begin{tabular}{lcr}
\toprule
\textbf{Layer type} & \textbf{Output shape} & \textbf{Parameters} \\
\midrule
InputLayer (input image) & (None, 4096) & 0 \\
InputLayer (target image) & (None, 4096) & 0 \\
InputLayer (operators) & (None, 64, 64, 8192) & 0 \\
Concatenate & (None, 8192) & 0 \\
Dense & (None, 128) & 1,048,704 \\
LeakyReLU & (None, 128) & 0 \\
Dense & (None, 128) & 16,512 \\
LeakyReLU & (None, 128) & 0 \\
Dense & (None, 64) & 8,256 \\
Dense & (None, 64) & 4,160 \\
\bottomrule
\label{tab:discriminator_model_summary}
\end{tabular}
\end{table}

Table \ref{tab:discriminator_model_summary} shows the specific layers used in the Discriminator model and the corresponding output shapes and trainable parameters in the case of $N=6$ qubits and $|M|=4^6 = 4,096$ measurement bases. This model does not differ significantly from the FCN model, only that another input is present which is the input image generated with the Generator model. Also, a Concatenate layer is used to merge the two input layers into a single tensor.

\subsection{Transformers Model}
Table~\ref{Transformer_table} shows the specific layers used in the Transformer model and the corresponding output shapes and trainable parameters in the case of $N=6$ qubits and $|M|=4^6 = 4,096$ measurement bases. The deviating layer compared to the previous models is the TransformerEncoder layer. Essentially, the encoder processes sequential data using self-attention to focus on the most significant parts of the data, potentially capturing features more efficiently. Since the decoder from the standard Transformer architecture is not required for QST, it is omitted, simplifying the model and improving efficiency. We set the number of attention heads to $8$ and the number of encoder layers to $4$.

\begin{table}[ht]
  \caption{Transformer model summary}
  \centering
  \begin{tabular}{@{}lcr@{}}
    \toprule
    Layer type              & Output shape           & Parameters        \\
    \midrule
    InputLayer              & (None, 4096)           & 0                  \\
    InputLayer              & (None, 64, 64, 8192)   & 0                  \\
    TransformerEncoder      & (1, 128, 128)          & 856,960           \\
    Flatten                 & (1, 16,384)           & 0                  \\
    Dense                   & (1, 8192)              & 134,225,920      \\
    Reshape                 & (1, 64, 64, 2)         & 0                  \\
    DensityMatrix           & (1, 64, 64)            & 0                  \\
    Expectation             & (1, 4096)              & 0                  \\
    \bottomrule
  \end{tabular}
  \label{Transformer_table}
\end{table}

\subsection{Spiking Variational Autoencoder Model}

\begin{table}[ht]
\caption{SVAE parameters summary.}
\centering
\begin{tabular}{lcr}
\toprule
\textbf{Parameter} & \textbf{Value} & \textbf{Description} \\
\midrule
Shots & 100,000 & Number of measurement shots \\
Beta & 0.819 & Hyperparameter for regularization \\
Number of Steps & 100 & Training steps per epoch \\
Number of Epochs & 5 & Total epochs for training \\
Learning Rate & $1 \times 10^{-3}$ & Initial learning rate \\
Number of Workers & 4 & For data loading \\
Shuffle & False & Data shuffling disabled \\
Input Size & $4 \cdot n$ & Proportional to the number of qubits ($n$) \\
Hidden Size & $20 \cdot n$ & Proportional to the number of qubits ($n$) \\
Output (Latent) Size & $2 \cdot 2^n$ & Proportional to the number of qubits ($n$) \\
Alpha & 1 & Scaling factor for loss terms \\
Model Recovery & False & No recovery of previous models \\
\bottomrule
\end{tabular}
\label{tab:svae_1}
\end{table}

\begin{table}[ht]
\caption{Parameters for training and validation across different numbers of qubits.}
\centering
\begin{tabular}{ccr}
\toprule
\textbf{Qubits} & \textbf{Batch Size for Training} & \textbf{Validation Samples} \\
\midrule
3 & 100 & 20,000 \\
4 & 300 & 100,000 \\
5 & 600 & $4^5 \cdot 500 = 512,000$ \\
6 & 600 & $4^6 \cdot 500 = 2,048,000$ \\
7 & 600 & $4^7 \cdot 500 = 8,192,000$ \\
8 & 1,000 & $4^8 \cdot 500 = 32,768,000$ \\
\bottomrule
\end{tabular}
\label{tab:svae_2}
\end{table}

The SVAE model relies on a set of key parameters that govern its training and validation processes, as summarized in Tables~\ref{tab:svae_1} and \ref{tab:svae_2}. Table~\ref{tab:svae_1} lists the common parameters applied across all SVAE model tests, such as the total number of shots ($100,000$), hyperparameters like beta ($0.819$), learning rate ($1 \times 10^{-3}$), and architectural choices that scale proportionally with the number of qubits, including input size ($4 \cdot n$), hidden size ($20 \cdot n$), and output (latent) size ($2 \cdot 2^{n}$). Other important parameters include the number of training steps per epoch ($100$), total epochs ($5$), number of data-loading workers ($4$), and disabling of data shuffling during training.

Table \ref{tab:svae_2} further details how training and validation are adapted for different numbers of qubits. The batch size for training gradually increases from $100$ for $3$ qubits up to $1,000$ for $8$ qubits, while the validation sample sizes grow exponentially with qubit number, following the pattern $4^{n} \times 500$, reaching over 32 million samples for $8$ qubits. This scaling reflects the combinatorial complexity of quantum state representations and ensures the model is validated with sufficient data to capture the increasing state space.

\end{appendices}


\bibliography{sn-bibliography}

\newpage

\end{document}